\documentclass[12pt,nohyper,letterpaper]{JHEP3}         
\usepackage{amssymb,amsfonts}

\usepackage{epsfig}

\def\be{\begin{eqnarray}}
\def\ee{\end{eqnarray}}
\newcommand{\nn}{\nonumber}
\newcommand\para{\paragraph{}}
\newcommand{\ft}[2]{{\textstyle\frac{#1}{#2}}}
\newcommand{\eqn}[1]{(\ref{#1})}

\def\Dslash{\,\,{\raise.15ex\hbox{/}\mkern-12mu D}}
\def\Dbarslash{\,\,{\raise.15ex\hbox{/}\mkern-12mu {\bar D}}}
\def\delslash{\,\,{\raise.15ex\hbox{/}\mkern-9mu \partial}}
\def\delbarslash{\,\,{\raise.15ex\hbox{/}\mkern-9mu {\bar\partial}}}
\def\pslash{\,\,{\raise.15ex\hbox{/}\mkern-9mu p}}
\def\calDslash{\,\,{\raise.15ex\hbox{/}\mkern-12mu {\cal D}}}

\def\lae{\mathrel{\mathop{\smash{\lower .5 ex \hbox{$\stackrel<\sim$}}}}}
\def\lae{\mathrel{\mathop{\smash{\lower .5 ex \hbox{$\stackrel>\sim$}}}}}

\newcommand\vphi{\!\vec{\,\phi}}

\title{The Partonic Nature of Instantons}
\author{Benjamin Collie and David Tong \\
Department of Applied Mathematics and Theoretical Physics, \\
University of Cambridge, UK\\
{\tt b.p.collie, d.tong@damtp.cam.ac.uk}}

\abstract{In both Yang-Mills theories and sigma models, instantons are endowed
with degrees of freedom associated to their scale size and orientation. It has
long been conjectured that these degrees of freedom have a dual interpretation as
the positions of partonic constituents of the instanton. These conjectures are
usually framed in $d=3+1$ and $d=1+1$ dimensions respectively where the
partons are supposed to be responsible for confinement and other strong
coupling phenomena. We revisit this partonic interpretation of instantons in the
context of $d=4+1$ and $d=2+1$ dimensions. Here the instantons are particle-like
solitons and the theories are non-renormalizable. We present
an explicit and calculable model in
$d=2+1$ dimensions where the single soliton in the ${\bf CP}^N$ sigma-model can be shown to be
a multi-particle state
whose partons are identified with the ultra-violet degrees of freedom which
render the theory well-defined at high energies. We introduce a number of methods
which reveal the partons inside the soliton, including deforming the sigma model and a dual
version of the Bogomolnyi equations. We conjecture that partons inside
Yang-Mills instantons hold the key to understanding the ultra-violet completion
of five-dimensional gauge theories.}

\begin{document}
\pagestyle{plain} \setcounter{page}{1}
\newcounter{bean}
\baselineskip16pt

\section{Introduction}

Solitons in field theory often have, in addition to their translational
degrees of freedom, a number
of further collective coordinates that arise from the action of internal symmetries.
Two prime examples of this are:
\begin{itemize}
\item Solitons in the ${\bf CP}^{N-1}$ sigma-model\footnote{These solitons
carry a bewildering number
of aliases. They are usually referred to as ``sigma-model lumps", sometimes
as  ``baby
skyrmions" and, in the condensed matter literature, simply as ``skyrmions".
They are closely related
to ``semi-local vortices". In the context of string theory
they are called ``worldsheet instantons".}: The single soliton has two
translational modes, a scaling mode
and $2N-3$ orientational modes arising from the $SU(N)$ global
symmetry. This gives $2N$ collective coordinates in total.
\item Yang Mills instantons: The instanton in $SU(N)$ gauge theory has 4
translational modes, a scaling mode and $4N-5$
orientational modes coming from large
$SU(N)$ gauge transformations. This gives $4N$ collective coordinates in total.
\end{itemize}
In both of these cases, all collective coordinates are Goldstone
modes arising from the underlying symmetries of the theory. However, it has long
been conjectured that, under certain circumstances, there
may be a different interpretation for these collective coordinates as the
positions of $N$ constituent objects which make up the soliton \cite{first}.
(More recent proposals along these lines include \cite{dm,ssz,z}).
These constituents have been christened with a variety of names over the years, from
the mundane ``fractional instanton" or ``instanton quark" to more flowery ``zindon" or ``quink". Throughout this paper, we err on the side of the mundane and refer to the instanton constituents
simply as ``partons".

\para
Conjectures about the partonic nature of instantons are usually framed in the context of
strongly coupled phenomena in $d=1+1$ dimensions (for sigma-models) and  $d=3+1$ dimensions (for
Yang-Mills theories). Such discussions typically hinge on the hope that some class of field
configurations dominates the path integral, even at strong coupling, resulting in a vacuum
which can be understood as a soup of correlated partons.

\para
In this paper we will
discuss the role of these partons in $d=2+1$ dimensions (for sigma-models) and $d=4+1$ dimensions
(for Yang-Mills theories). In these dimensions, the instanton solutions are particle-like solitons.
The theories are weakly coupled in the infra-red, but non-renormalizable and require completion in the ultra-violet (UV). The premise of this paper is that, in some situations, the partons provide the degrees
of freedom that form this UV completion.

\EPSFIGURE{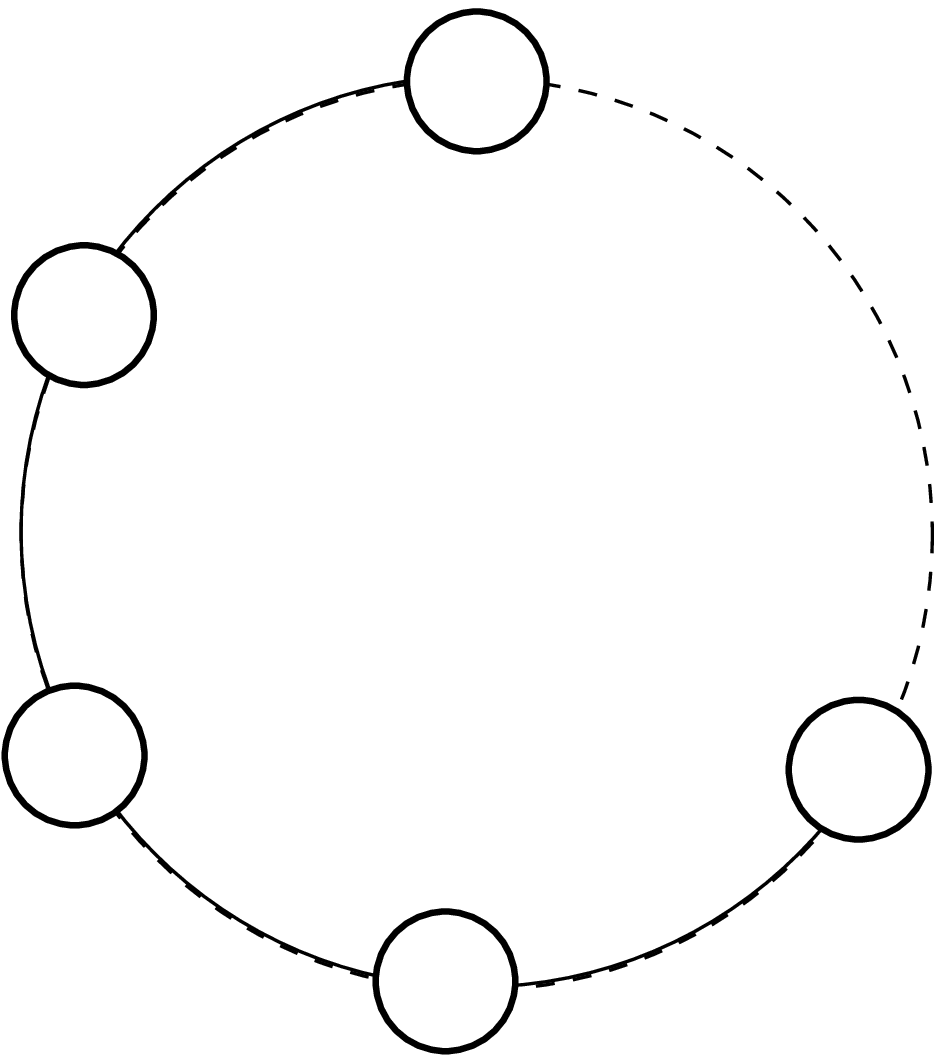,height=110pt}{}
\para
The main purpose of this paper is to provide a detailed example where the partonic
nature of solitons is explicit, and under analytic control. The example
that we offer is a supersymmetric gauge theory in $d=2+1$ dimensions which flows,
in the infra-red, to a variant of the ${\bf CP}^{N-1}$ sigma-model. The field
content of the gauge theory is shown in the quiver diagram: it consists of a $U(1)^{N}$
gauge group, together with $N$ matter fields carrying charge $(+1,-1)$ under consecutive $U(1)$
factors.

\para
The sigma-model of interest arises as the Coulomb branch of this theory. This means that
after integrating out the matter multiplets, and dualizing the photons, the low-energy
dynamics of the gauge theory is captured by a suitable variant of the ${\bf CP}^{N-1}$ sigma-model
\cite{is}. (To be precise, the Coulomb branch is the cotangent bundle $T^\star {\bf CP}^{N-1}$).

\para
Although we integrate out the charged matter to arrive at the sigma-model description,
we have not lost all trace of it. Its memory remains in the guise of the soliton. A single
soliton can be shown to be a multi-particle state, formed from the matter multiplets
that have been integrated out. Our goal in this paper is to answer the inverse question:
``Given access to the low-energy sigma-model, what can we learn about the UV gauge theory
through a study of the soliton?"  Surprisingly, we show that  the properties
of the soliton allows us to  reconstruct the microscopic quantum numbers of the partons,
details which one would imagine had been swept away by the winds of the renormalization group.

\para
We introduce a number of methods that demonstrate the partonic nature of the sigma-model
lumps. Perhaps the cleanest is to study the instantons in a deformation of the sigma-model,
in which a single soliton solution decomposes into $N$ constituents and allows us to graphically
demonstrate the existence of partons. We further show that the soliton
equations can be re-interpreted as the equations of electrically charged point-like sources.
In this reformulation, the scale and orientation moduli of the soliton can be explicitly seen
to be the positions of $N$ partons carrying the quantum numbers dictated by the microscopic
quiver theory. Moreover, this provides insight into the manner
in which fundamental fields morph into solitons. Finally, we also describe the relationship
between our partons and calorons which arise when the theory is compactified on a circle.

\para
All the work presented in this paper was undertaken with an eye to the harder problem of partons
in Yang-Mills instantons. We hope to address this question in a future publication and, for now, limit
ourselves only a few relevant observations which form the content of  Section 4.

\section{Partons in the ${\bf CP}^1$ Sigma Model}

In this section we discuss solitons in the ${\bf CP}^1$ sigma model. We describe several
methods which reveal the two partons sitting inside each soliton. All  of these methods can
be generalized to the  ${\bf CP}^{N-1}$ sigma model at the expense of
more cumbersome notation which is postponed until Section \ref{cpn}.

\para
We will consider a sigma model with target space $T^\star{\bf CP}^1$, that is the cotangent
bundle of ${\bf CP}^1$. This has the same soliton spectrum as ${\bf CP}^1$ but admits
an extension to a theory with ${\cal N}=4$ supersymmetry. (This means eight supercharges in
$d=2+1$ dimensions). The sigma model is non-renormalizable and requires a UV completion.
Following \cite{is}, we realise this UV completion by constructing the sigma model
as the Coulomb branch of a gauge theory. We
start by reviewing this construction.

\subsection{A UV Completion of the Sigma Model}

${\cal N}=4$ theories in $d=2+1$ dimensions have a global $SU(2)_R\times SU(2)_N$ R-symmetry.
The theories we consider in this paper are  built from two standard multiplets:
\begin{itemize}
\item The vector multiplet contains a gauge field $A_\mu$ and three real scalar fields $\vphi$
transforming in the $({\bf 1},{\bf 3})$ representation of the R-symmetry group. There are also four
Majorana fermions transforming as $({\bf 2},\bar{\bf 2})$.
\item The hypermultiplet contains a doublet of  complex scalar fields
$Q=(q,\tilde{q}^\dagger)$, transforming as $({\bf 2},{\bf 1})$ under the R-symmetry. The four Majorana fermions transform as $({\bf 1},{\bf 2})$.
\end{itemize}
Our theory consists of a single $U(1)$ vector multiplet coupled to two hypermultiplets
$Q_1$ and $Q_2$ with charge $+1$ and $-1$ respectively. The bosonic part of the Lagrangian is
given by,
\be -{\cal L} &=& \frac{1}{4e^2}F_{\mu\nu}F^{\mu\nu} + \frac{1}{2e^2}(\partial_\mu \phi)^2
+\sum_{i=1}^2 \left(|{\cal D}_\mu q_i|^2 + |{\cal D}_\mu \tilde{q}_i|^2 \right)
\nn\\ && + (\vec{m}+\!\vec{\,\phi})^2(|q_1|^2
+|\tilde{q}_1|)^2 + (\vec{m}-\vphi)^2 (|q_2|^2+|\tilde{q}_2|^2)\nn \\
 &&+\frac{e^2}{2}(|q_1|^2-|q_2|^2-|\tilde{q}_1|^2+|\tilde{q}_2|^2)^2
 +2e^2|\tilde{q}_1q_1-\tilde{q}_2q_2|^2\ .\label{lag}
\ee
Here $e^2$ is the gauge coupling constant,while $\vec{m}$ is a triplet of mass
parameters. The vector multiplet is massless while, for generic values of $\vphi$, the
hypermultiplets are massive. The theory has a single flavour symmetry, $U(1)_F$, under
which the hypermultiplets $Q_1$ and $Q_2$ both have charge $+1$.

\para
We are interested in the low-energy effective action for the
vector multiplet. This is best described after first dualizing the photon in favour of
a periodic scalar field, $\sigma \in [0,2\pi)$, defined by
\be F_{\mu\nu} = \frac{e^2}{2\pi}\,\epsilon_{\mu\nu\rho}\partial^\rho \sigma\label{dual}\ee
Written in the dual variables the theory has a further global symmetry, usually denoted as $U(1)_J$,
which acts by shifting $\sigma$. In non-Abelian theories this symmetry is typically broken by
instanton effects, but in our Abelian theory it remains exact.

\para
After integrating out the hypermultiplets at one-loop, the low-energy effective action is
given by a sigma model on the Coulomb branch \cite{is},
\be -{\cal L} = \frac{1}{2}H(\phi)\,(\partial_\mu \vphi)^2  + \frac{1}{8\pi^2} H(\phi)^{-1}(\partial_\mu \sigma + \vec{\omega}\cdot
\partial_\mu \vphi)^2\label{tn}\ee
Here the function $H(\phi)$ can be thought of as the renormalized gauge coupling, receiving
contributions from each of the two hypermultiplets
\be
H=\frac{1}{e^2}+\frac{1}{4\pi|\vec{m}+\vphi|}+\frac{1}{4\pi |\vec{m}-\vphi|}\label{h}\ee
The factor of $4\pi$ in this expression is usually neglected, but
arises from an explicit one-loop computation as shown in \cite{dkmtv}. This normalization
will prove important later in our discussion. The connection $\vec{\omega}$ in \eqn{tn} is
defined by $\vec{\nabla} H = \vec{\nabla}\times \vec{\omega}$.

\para
The one-loop effective action \eqn{tn} defines a sigma-model with a hyperK\"ahler metric
on two-centered Taub-NUT space. This hyperK\"ahler structure is required by supersymmetry
and is sufficient to ensure that there are no further corrections to the action: the one-loop result
is exact \cite{is}. In particular, it holds even in the strong coupling limit
$e^2\rightarrow \infty$. Here something special
happens: the $U(1)_J$ isometry is enhanced to $SU(2)_J$ and the metric \eqn{tn} becomes
the Eguchi-Hanson metric on $T^\star{\bf CP}^1$ \cite{eh}.

\subsubsection*{The ${\bf CP}^1$ submanifold}

We will be interested in the ${\bf CP}^1$ submanifold that is the zero section of
$T^\star{\bf CP}^1$. It is also sometimes known as the ``bolt". To define it, we
choose for simplicity $\vec{m}=(0,0,m)$. The bolt is then defined as the submanifold
with $\vphi=(0,0,\phi)$ and $\phi  \in [-m,m]$.
The metric on the bolt is given by
\be ds^2 = H(\phi)\, d\phi^2 + \frac{1}{4\pi^2}H(\phi)^{-1}\,d\sigma^2\nn\ee
with
\be H(\phi) = \frac{1}{e^2} + \frac{m}{2\pi(m^2-\phi^2)}\label{H}\ee
For finite $e^2$, this is the metric on a squashed sphere written in ``toric" coordinates.
When $e^2\rightarrow \infty$, it becomes the metric on the round sphere with $SU(2)_J$ isometry.
To see this explicitly, we define the complex coordinate on the Riemann sphere,
\be R = \sqrt{\frac{m-\phi}{m+\phi}}\,e^{i\sigma}\label{r}\ee
in terms of which the metric, in the $e^2\rightarrow\infty$ limit, takes the familiar form,
\be ds^2 = \frac{2m}{\pi}\frac{dR\, d\bar{R}}{(1+|R|^2)^2}\label{round}\ee

\subsection{Solitons and Their Microscopic Interpretation}

The low-energy sigma model has solitons. These solitons are BPS only if we
take the vacuum to lie on the ${\bf CP}^1$ bolt defined above. (In fact, if this
is not the case, the soliton profile does not have a well-defined asymptotic limit). In
this section we study the properties of the soliton and identify  this object
in the microscopic gauge theory.

\para
Let us first determine the mass of the soliton. It is related to the size of the ${\bf CP}^1$
(and this is the reason that the factor of $1/4\pi$ was important in \eqn{h}). It is
a simple matter to write down the lump equations in terms of the $\phi$ and $\sigma$ fields.
The energy functional for static configurations can be written as:
\be {\cal E} &= &\int d^2x\ \frac{1}{2}H\,\partial_\alpha\phi\partial_\alpha\phi +
\frac{1}{8\pi^2}H^{-1}\partial_\alpha\sigma\partial_\alpha\sigma \nn\\
&=& \int d^2x\ \frac{1}{8\pi^2}H^{-1}(2\pi H\,\partial_\alpha \phi \mp
\epsilon_{\alpha\beta}\partial_\beta\sigma)^2
\pm \frac{1}{2\pi}\epsilon_{\alpha\beta}\partial_\alpha\phi\partial_\beta\sigma.
\label{square}\ee
The Bogomolnyi equations can be found sitting
within the total square: they are
\be 2\pi H(\phi)\, \partial_\alpha\phi = \epsilon_{\alpha\beta}\,\partial_\beta \sigma\label{bog}\ee
where the function $H(\phi)$ is given in \eqn{H}. When these equations are
satisfied, the energy is given by the last term in \eqn{square}
which we recognize as the topological charge. It counts the winding
of the configuration, weighted by the area of the (squashed) sphere.
Recalling that $\phi\in[-m,m]$ and $\sigma \in [0,2\pi)$, this area is given by $4\pi m$.
The mass of the BPS lump, given by the lowest energy configuration with
unit winding number, is
\be M_{\rm lump} = 2m\nn\ee
So what is this object in the microscopic gauge theory? We're looking for a BPS state
with mass $2m$. There is only one candidate: the soliton corresponds to a two particle
state $Q_1Q_2$ constructed from the hypermultiplets. This state is neutral under the
$U(1)$ gauge symmetry, but charged under the $U(1)_F$ flavour symmetry.
The flavour charge has morphed into the topological charge at low energies.

\para
The soliton is BPS only for vacua that lie on the ${\bf CP}^1$ bolt. But
this is also true of the state $Q_1Q_2$: the requirement that it is BPS is that the
two mass-vectors $\vec{m}\pm \vphi$ are parallel. This holds when $\vphi$ and $\vec{m}$
lie parallel, with $|\vphi| \leq |\vec{m}|$. At low energies this descends to the requirement
that we lie on the bolt.

\para
This is quite cute. We integrated out the hypermultiplets and might have expected that
we'd lost them for good. But, in fact, they re-appear in the low-energy effective
action as solitons. It is somewhat reminiscent of the manner in which baryons appear as
skyrmions in the chiral Lagrangian. The identification of the soliton with a multi-particle
state  was first made in the context  of mirror symmetry
as particle/vortex duality \cite{ahiss} and was elaborated upon further in \cite{3d2d,kap}. In
the rest of this section, we will study the implications of this identification in more detail.

\para
The first question that we should answer is: why are the partons bound to form pairs within
the soliton? The reason is that, in three dimensions,
the $1/r$ fall-off of the electric field ensures that any state charged under a local current
has logarithmically divergent mass. There is a similar IR divergence from the massless $\phi$ field.
This means that on the Coulomb branch, where the gauge symmetry is unbroken, all finite
mass states are associated to gauge invariant operators. In our theory the only such
BPS operator is the dipole $Q_1Q_2$.

\para
Although the infra-red divergence requires that the partons are bound together, there is no
static force between them. This is manifested in the solitonic description by the existence of four
collective coordinates. Two simply give the center of mass of the soliton, $Z$. The remaining two
correspond to a scale size $\rho$ and an orientation collective coordinate $\theta$. In the
limit $e^2\rightarrow \infty$, where the target space becomes the round sphere, $\theta$ is a
Goldstone mode arising from the action of $U(1)_J^\prime \subset SU(2)_J$ on the soliton. This
$U(1)_J^\prime$ is defined by the requirement that it leaves the vacuum invariant and, in general, does
not coincide with $U(1)_J$. (We will explain an exception to this statement below). As we will describe in detail, when  $e^2$  is finite and the
target space sphere is squashed, $\theta$ is not in general associated with a Goldstone mode.

\subsection{How to Tell if Your Soliton Contains Partons}
\label{2.3}

The microscopic interpretation of the soliton is as a dipole of charged hypermultiplets.
We would like to ask what memory the soliton has of its microscopic origins. In other words,
suppose that we have access only to
low-energy information captured within the sigma model: what would we be able to
say about the hypermultiplets that we have integrated out?  Here we offer a number
of  methods which reveal the partonic nature of the soliton. Throughout this section
we work in the
vacuum $\langle\phi\rangle=0$ where
the two partons have equal mass $m$. (We will relax this condition in Section \ref{caloron}).

\subsubsection{Deforming the Sigma Model}
\label{2.2.4}

\EPSFIGURE{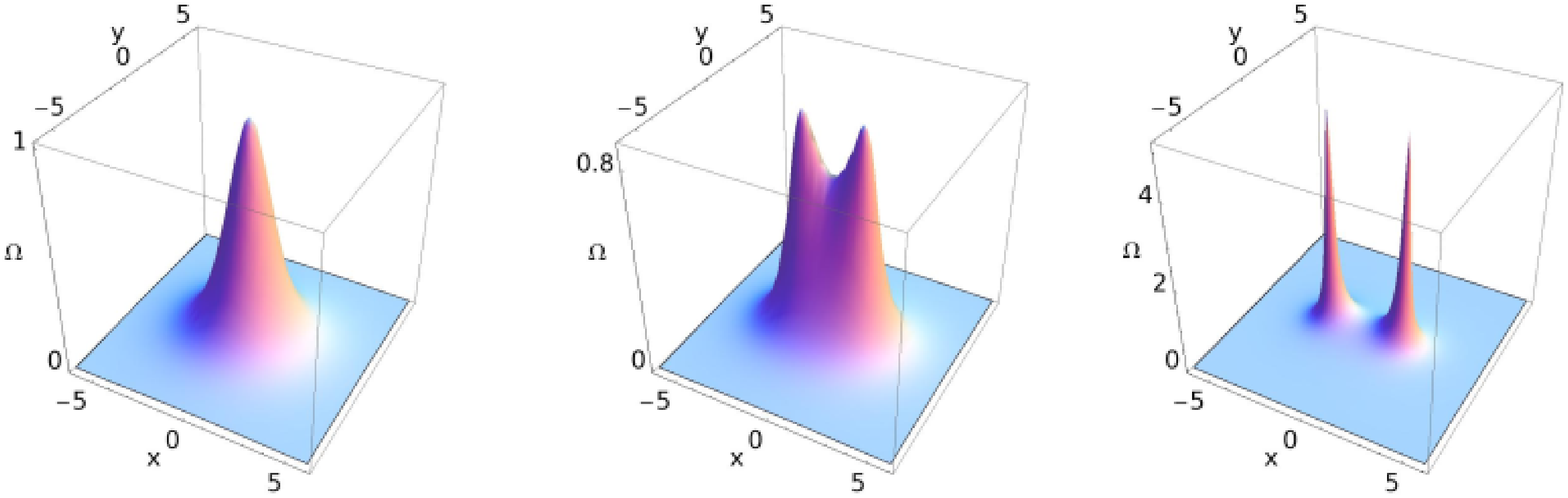,width=15cm}{Energy density for a single  BPS soliton in ${\bf CP}^1$
with $m/e^2 = 0$, $1/2$, $3/2$.}{\label{cp1fig}}

The simplest and most explicit method which reveals the partons is to look
at the single soliton solution in the deformed ${\bf CP}^1$ sigma-model. The deformation
that we have in mind occurs naturally in our UV theory: it is the
squashed target space with $H(\phi)$ given by \eqn{H} with finite $e^2$.

\para
When $m/e^2=0$, and the target space inherits the round metric, a plot
of the energy density reveals no hint of the microscopic structure.
It simply gives a blurred lump of size $\rho$ as shown in the first plot of Figure 2.
The remaining two plots in the figure show the energy density of the BPS soliton as $m/e^2$ increases
and the target space is squashed into the shape of a rugby ball\footnote{Mathematica notebooks 
for all figures presented in this paper can be downloaded from 
http://www.damtp.cam.ac.uk/user/tong/parton.html\ .}. The topological charge becomes
localized around the tips at $\phi = \pm m$. In space, we see that the energy density becomes
localized in two equal peaks, each with ${\cal E} = m$, sitting at positions $z_+$ and $z_-$,
separated by a distance $|z_+-z_-|=2\rho$.

\para
In the limit $m/e^2\rightarrow \infty$, the parton configuration in the figure is reminiscent of
the ``meron" configuration described in \cite{gross}. A single parton has topological charge $1/2$, as does a single meron.
However, there are also differences. Our configuration is a  solution to the equations of motion
on the squashed sphere; the meron configuration is a singular solution to equations on the round sphere. This difference
becomes most manifest when we look to generalize the discussion to the ${\bf CP}^{N-1}$ sigma-model.
The meron continues to have topological change $1/2$ in this context while, as we will see in
the following section, deforming the metric on ${\bf CP}^{N-1}$ causes the lump to decompose
into $N$ partons with topological charge $1/N$. A similar mechanism for revealing the partonic structure
of sigma-model lumps was independently found in \cite{competition}\footnote{Note added: After the
first version of this paper appeared on the arXiv, it was pointed out to us that
 equation (4.26) of \cite{hidden} also reveals the energy density of a sigma-model lump with
multiple peaks.
In this case, the reason for the partonic behaviour appears to be rather different, resulting from  the singular nature of the target space which, in turn, gives rise to a singular energy density.}.

%
%
%

\subsubsection{Parton Quantum Numbers}
\label{2.2.2}

We saw above that deforming the sigma model target space dramatically reveals the partonic
nature of the soliton. But suppose that we work in the $e^2\rightarrow \infty$ limit
where the target space is round and the energy density is merely a smeared blob. Is it
still possible to disentangle the partonic structure? The answer, as we shall see, is yes.

\para
The first hint at a partonic structure was described long ago in \cite{first} and arises
from simply looking at the solution in different variables. In the $e^2\rightarrow
\infty$ limit, the profile of a single soliton is given by
\be \phi  = m\rho \,\frac{e^{-i\theta} (z-Z) + e^{i\theta}(\bar{z}-\bar{Z})}{|z-Z|^2+\rho^2}
\nn\ee
The four collective coordinates of the solution are $Z\in{\bf C}$, the centre of mass,  $\rho\in {\bf R}^+$, the scale size of the soliton, and
$\theta\in[0,2\pi)$, a Goldstone mode arising from the $U(1)^\prime_J \subset SU(2)_J$
flavour symmetry which is left unbroken by the choice of vacuum.

\para
However the fact that the collective coordinates can be re-interpreted as the positions of partons
is clear if we rewrite the soliton solution using the coordinate $R$ on target space, defined
in \eqn{r}.
Then the single soliton solution takes the form,
\be R = \frac{z-z_+}{z-z_-}\label{1lump}\ee
with $Z=\ft12(z_-+z_+)$ and $\rho e^{i\theta} = \ft12(z_--z_+)$.
The collective coordinates $z_+$ and $z_-$ reveal the positions of the partons. Indeed, at $z= z_{\pm}$ we have $\phi = \pm m$, so at each of these locations one of the hypermultiplets of our microscopic theory becomes massless.
We will shortly see that the soliton profile near these points reveals that the configuration carries the correct electric charge.

\para
There is a simple generalization of the single soliton solution \eqn{1lump} to a $k$ soliton solution,
\be R = \prod_{n=1}^k\frac{z-z_+^n}{z-z_-^n}\label{or2}\ee
where $\{z_+^n\}$ and $\{z_-^n\}$ denote the positions of the hypermultiplet excitations $Q_1$ and $Q_2$
respectively.  The fact that the $k$-lump solution is determined by the positions of two sets of $k$
points on the plane has long been taken as  evidence for the partonic nature of the soliton
\cite{first}. This fact is explicitly realised in the three dimensional
gauge theory construction.

\subsubsection*{Dual Bogomolnyi Equations}

The partons $Q_1$ and $Q_2$ in the microscopic model carry electric charge $\pm 1$
under the $U(1)$ gauge field. They also carry flavour charge $+1$ under $U(1)_F$. This
flavor symmetry descends to the topological charge of the sigma model. But it is also
possible to reconstruct the electric charge of the partons from the lump solutions.

\para
To do this, we rewrite the soliton equations \eqn{bog} in terms of dual
variables. We begin by inverting the duality transformation \eqn{dual}, this time
with the low-energy gauge field $F_{\mu\nu}$ defined in terms of the renormalized
gauge coupling,
\be F_{\mu\nu} = \frac{H^{-1}}{2\pi}\,\epsilon_{\mu\nu\rho}\partial^\rho \sigma\label{d2}\ee
In these variables, the Bogomolnyi equation \eqn{bog} simply relates the electric field to
the variation of $\phi$,
\be \partial_\alpha \phi = F_{0\alpha}\label{dbog}\ee
Although these equations are merely a re-writing of the Bogomolnyi equations, they do {\it not}
have smooth solutions corresponding to solitons. Instead, in order to reproduce the soliton
profiles, we must introduce point-like sources. This is to be expected in an electrical
formulation of the theory.

\para
The transformation \eqn{d2} ensures that the electric field is divergence free {\it except}
at points where $\sigma$ is ill-defined. It is simple to see where these points lie from the expressions
\eqn{r} and \eqn{or2}: they are at $z=z_+^n$ and $z=z_-^n$. Since $\sigma$ has non-trivial winding
around each of these points, they act as sources for the electric field.  In particular, we note from \eqn{or2} that $\sigma$ increases by $2\pi$ if we complete an anticlockwise circuit around $z_+^n$, and $\sigma$ decreases by $2\pi$ if we complete an anticlockwise circuit around $z_-^n$.  Hence for an arbitrary closed loop $\mathcal{C}$ which avoids the points $z_\pm^n$ and encloses a region $\mathcal{S}$, we have
\be
\int_\mathcal{C} dx_\alpha\ \partial_\alpha \sigma =
2\pi \int_\mathcal{S} dS\ \sum_n \left[ \delta(z-z^n_+) - \delta(z-z^n_-) \right]
\label{delta}\ee
We can rewrite the left-hand-side of this equation  using Stokes' theorem and the duality \eqn{d2}.
Since the resulting equation holds for arbitrary regions $\mathcal{S}$, we can equate the integrands
to find,
\be \partial_\alpha(HF_{0\alpha}) = \sum_n\ \left[ \delta(z-z^n_+) - \delta(z-z^n_-)\right].\label{source}\ee
This is a rather novel method of viewing soliton collective coordinates as sources. For each value
of $\{z^n_+\}$ and $\{z^n_-\}$, there is a unique solution to \eqn{dbog} and \eqn{source}.
This determines a point on the soliton moduli space.
Typically, the electrically charged particles in a theory arise as fundamental excitations, while magnetically charged objects are associated to solitons. This simple model in three dimensions
provides an example where we can swap between these two descriptions with ease.

\subsubsection{The Force Between Solitons}

The dual Bogomolnyi equations above reveal that the partons carry electric charge,
and we already mentioned that this is responsible for the logarithmic confinement of the
partons with the soliton.  Although there is no static force between the partons, the logarithmic
infra-red divergence suffered by any individual parton reappears once we ask them to move.
It is well known that the moduli space metric for sigma-model lumps has logarithmic infra-red
divergences \cite{ward}. For ${\bf CP}^1$ lumps, there is just a single divergent mode that arises
from the long-range tail of $R$,
\be \dot{R} \sim \sum_n\frac{\dot{z}_+^n-\dot{z}_-^n}{z}+\ldots
\nn\ee
The kinetic terms are finite only if the sum $\sum_n (z_+^n-z_-^n)$ is constant. This is precisely
the condition that the sum of the dipole moments is unchanged, as expected from the microscopic
theory.

\para
While solitons have only velocity-dependent forces, there is an attractive force between
a lump and an anti-lump. This calculation was performed many years ago and provides
yet another method to illuminate the partonic structure of the soliton \cite{gross}.
One starts by constructing a configuration describing a well-separated lump and anti-lump.
The lump has size $\rho$ and orientation $\theta$ and is placed at the origin. The anti-lump
has size $\bar{\rho}$ and orientation $\bar{\theta}$ and centre of mass position $z=r e^{i\chi}$,
where $r\gg \rho,\bar{\rho}$. The interaction energy between the two objects is then computed
to be \cite{gross}
\EPSFIGURE{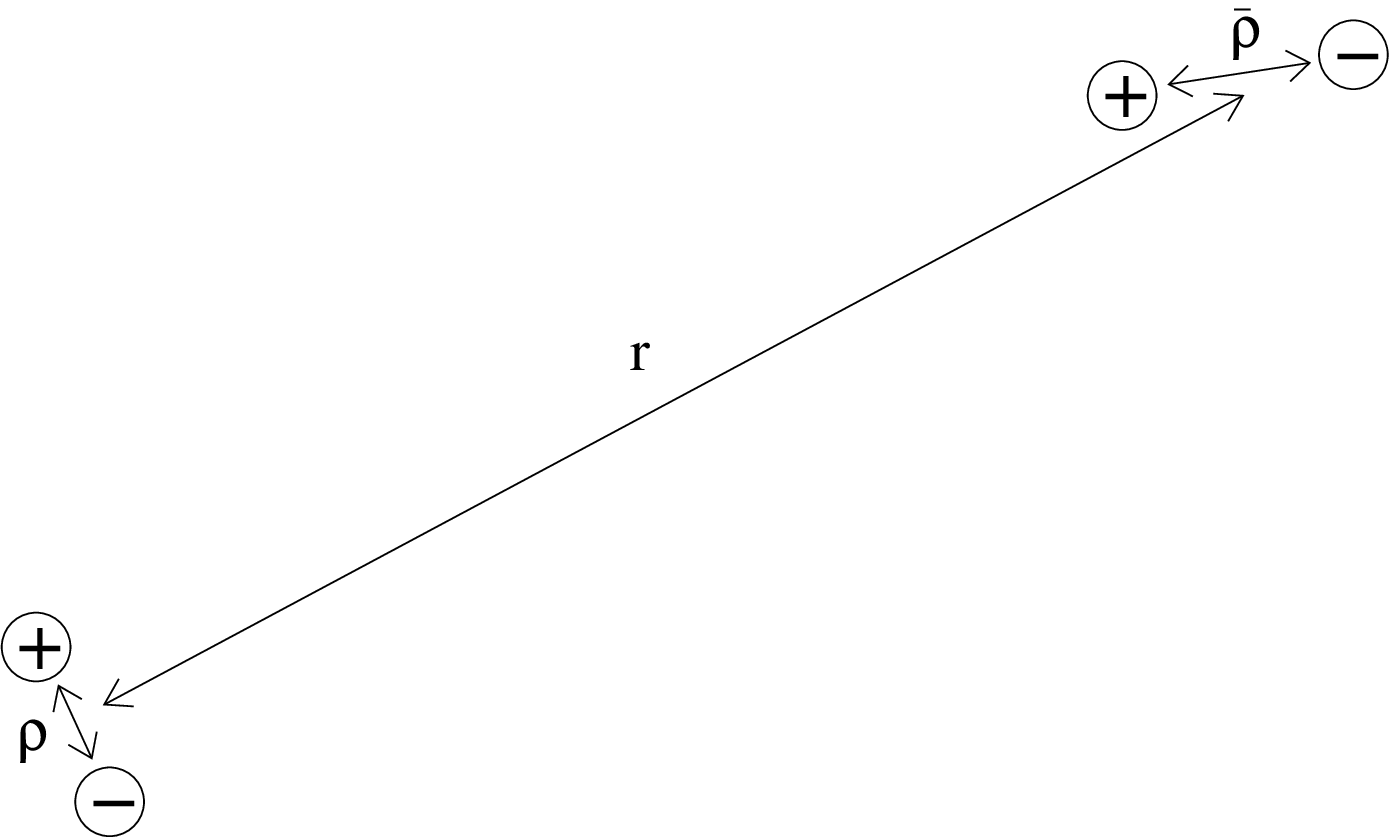,height=80pt}{}
\be V_{\rm int} = - 4m\,\frac{\rho\bar{\rho}}{r^2}\,\cos\left(\theta+\bar{\theta}-2\chi\right)\nn\ee
This is precisely the interaction energy of two dipoles on the plane, with orientation
$\rho e^{i\theta}$ and $\bar{\rho}e^{i\bar{\theta}}$, separated by $re^{i\chi}$.  Indeed,
it can be put in slightly more familiar form if we define $\vec{\rho}_1=\rho(\cos\theta,
\sin\theta)$, $\vec{\rho}_2=\bar{\rho}(\cos\bar{\theta},\sin\bar{\theta})$ and
$\vec{r}=r(\cos\chi,\sin\chi)$. Then the interaction potential can be written as the
dipole interaction,
\be V_{\rm int} = -\frac{4m}{r^2}\left(2(\vec{\rho}_1\cdot\hat{\vec{r}})
(\vec{\rho}_2\cdot\hat{\vec{r}}) - \vec{\rho}_1\cdot\vec{\rho}_2\right)\nn\ee
It is
remarkable that this inter-soliton force captures the partonic structure in such a
clean fashion.

\para
Finally, it's worth mentioning another famous calculation which, while not directly
relevant to the present discussion, also reveals the
partonic nature of instantons in the ${\bf CP}^{1}$ sigma-model. This is the computation
of the determinants around the background of multiple lumps for the theory in $d=1+1$ dimensions
\cite{frolov,berg}. This computation reveals a dipole-like structure for these objects even when
viewed as instantons localized in Euclidean spacetime.

\subsection{Relationship to Calorons}
\label{caloron}

There is another context in which it is known that the sigma-model lump decomposes into partons,
known as {\it calorons}. To achieve this, one compactifies the theory on a spatial circle
of radius $L$. After deforming the theory in a suitable manner (to be described below),
the lump in the ${\bf CP}^{N-1}$ sigma-model can be shown to decompose into $N$ domain
walls \cite{me}. (This phenomenon was further discussed in \cite{eto} and recently
rediscovered in \cite{bruck1,harl,bruck2}).
Importantly, a similar phenomena also occurs for Yang-Mills instantons
compactified on a circle \cite{leeyi,vanbaal}. For this reason, we spend some time in
this section describing this phenomenon in our sigma model and examining the relationship
between the calorons and the partons.

\subsubsection*{Changing the Vacuum}

Until now,  we studied the solitons around the vacuum $\langle\phi\rangle=0$. From the perspective
of the UV gauge theory, this ensures that the partons have equal bare mass, $m$. In order
to understand the relationship to calorons, we will first look at the behaviour of the
solitons as we change the vacuum.

\para
As we vary the vacuum, the microscopic masses of the partons change:  they become
$m\pm \langle\phi\rangle$. The energy density for a single ${\bf CP}^1$
soliton is shown in Figure 4 as we vary the vacuum. It is clear that the energy in each
spike changes, although the difference in the heights of the spike is not linear in
$\langle\phi\rangle$ as one might naively expect from the classical theory. It appears that much
of the energy density is dispersed in the field between the solitons. It may be interesting to
explore this further, although we shall not do so here.

\EPSFIGURE{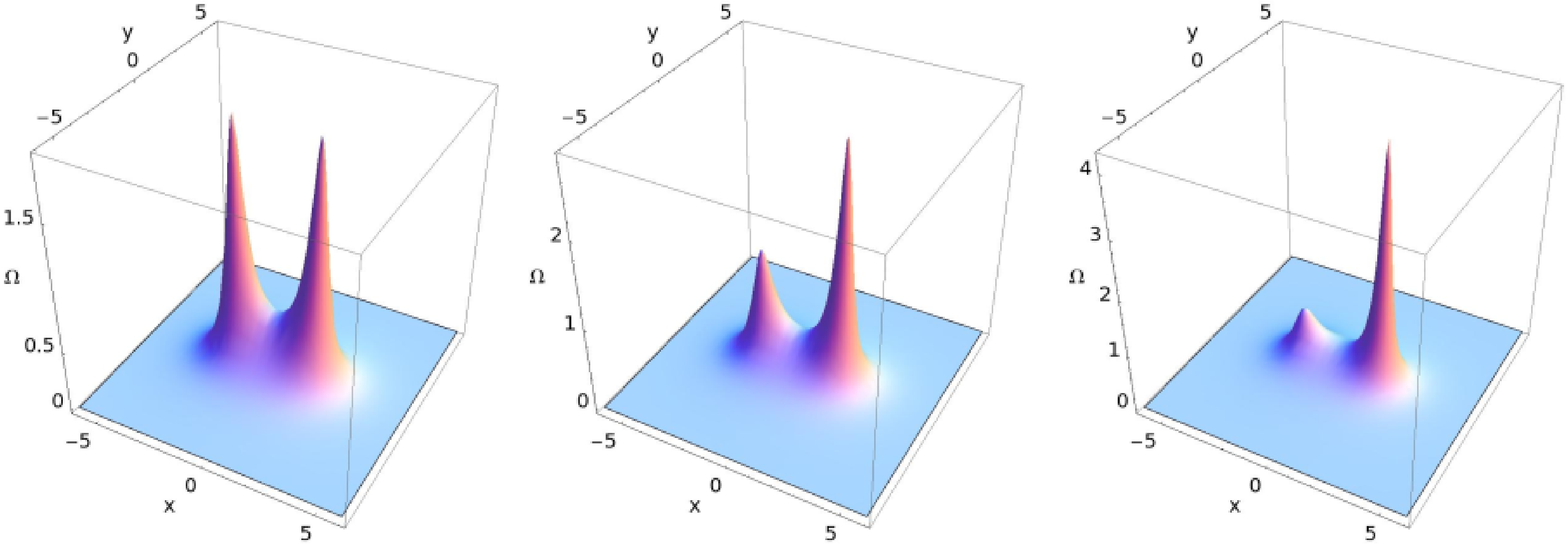,width=15cm}{A single ${\bf CP}^1$ soliton with $m/e^2 = 1$ as the vacuum varies from
$\langle{\phi}\rangle=0$
to $\langle\phi\rangle=0.1m$ to $\langle\phi\rangle=0.2m$.}
\para
There is one key feature that will be important in what follows. As
$\langle\phi\rangle \rightarrow \pm m$, only a single energy spike survives. In this
limit, one of the hypermultiplets in the microscopic theory becomes massless. It
is notable that the low-energy dynamics doesn't notice this fact. Typically, low-energy
effective theories become singular when further fields become massless. The reason
that this doesn't happen in three dimensions is due to the infinite energy contained in
the long range electric field that accompanies any charged state.
This ensures that even though the mass in the Lagrangian
vanishes, there are no extra massless charged excitations.

\para
In the context of the soliton, we learn that
the partonic description is really {\it not} an accurate reflection of the physics
when $\langle\phi\rangle = \pm m$. Instead, varying the scale size $\rho$, and the orientation
$\theta$ of the lump
in the deformed sigma-model does exactly what it says on the tin: it changes the scale size and
orientation. Microscopically
the scale size arises from exciting a cloud associated to the massless fields. Moreover, in this
limit the orientation mode $\theta$ is a Goldstone boson. This is because
the symmetry $U(1)_J^\prime$ that preserves the vacuum coincides with $U(1)_J$ and acts on the
soliton even when  $m/e^2\neq 0$.

\subsubsection*{Introducing a Potential}

Before we move on to describe the calorons in this model, it is useful to first recollect
what happens when we introduce a potential in the low-energy dynamics. This can be induced
in the microscopic theory by a Fayet-Iliopoulos parameter, $\zeta > 0$, after which the potential
terms in \eqn{lag} become
\be V &=&  (\vec{m}+\!\vec{\,\phi})^2(|q_1|^2
+|\tilde{q}_1|)^2 + (\vec{m}-\vphi)^2 (|q_2|^2+|\tilde{q}_2|^2)\nn \\
 &&+\frac{e^2}{2}(|q_1|^2-|q_2|^2-|\tilde{q}_1|^2+|\tilde{q}_2|^2-\zeta)^2
 +2e^2|\tilde{q}_1q_1-\tilde{q}_2q_2|^2\ .\label{newpot}\ee
This theory no longer has a moduli space of vacua, but rather two isolated vacua given
by $\vec{\phi}=-\vec{m}$, $|{q}_1|^2=\zeta$ and $\vec{\phi}=+\vec{m}$, $|\tilde{q}_2|^2=\zeta$.
Upon integrating out the hypermultiplets, this is reflected in our low-energy
description on the Coulomb branch (which is now strictly valid only for $m\gg \zeta$) by the
presence of the potential,
\be V = \frac{1}{2}\zeta^2H(\phi)^{-1}\label{zpot}\ee
The minima of this potential lie at $\phi=\pm m$. As described above, before we turned on this
potential, the partonic interpretation of the soliton was already rather different in these vacua.
After turning on the potential, the effect on the soliton is even more dramatic: it shrinks to
the singular solution with vanishing scale size $\rho=0$. This behaviour can be understood from the
microscopic theory. As can be seen from \eqn{newpot}, the presence of the FI parameter causes
the massless hypermultiplet to condense in the vacuum. This screens the massless cloud which provided
the non-zero size $\rho$ of the soliton.

\subsubsection*{Calorons}

We are now in a position to describe the emergence of calorons. We first compactify the spatial
direction $x^2$ on a circle of radius $L$. The reduced Lorentz symmetry allows for the addition
of one further interaction: a theta term $(\theta/4\pi^2 L) F_{01}$. The theta angle sits in a supermultiplet with the FI parameter $\zeta$ and, as we now explain, induces a potential
similar to \eqn{zpot}. Dualizing the photon in the presence of the theta term means that
\eqn{dual} becomes $(2\pi/e^2) F_{\mu\nu} = \epsilon_{\mu\nu\rho}\partial^\rho\sigma
- (\theta/2\pi) \epsilon_{\mu\nu 2}$ and the kinetic terms for the dual photon are
given by
\be \frac{e^2}{8\pi^2}\left(\partial_0\sigma^2-\partial_1\sigma^2 -(\partial_2\sigma
-\theta/2\pi L)^2\right)\nn\ee
Upon integrating out the hypermultiplets, $e^2$ is again renormalized by $H^{-1}(\phi)$ given in
\eqn{h}. The $\theta$ term in the action is then seen to generate a potential term. The effect
of the $\theta$ term can be mitigated if $\sigma$ winds $n\in {\bf Z}$ times around the compact
circle, so the potential is  given by
\be
V=
\left(\frac{\theta}{2\pi} - n\right)^2
\frac{H^{-1}(\phi)}{8\pi^2 L^2}
\nn\ee
The minima are again at $\phi=\pm m$. The physics here is similar to that of the FI
parameter. The $\theta$ angle induces a background electric field in two dimensions
\cite{coleman}. This can be screened by a condensation of charged scalars which can
only occur at $\phi=\pm m$ where these scalars are massless.

\EPSFIGURE{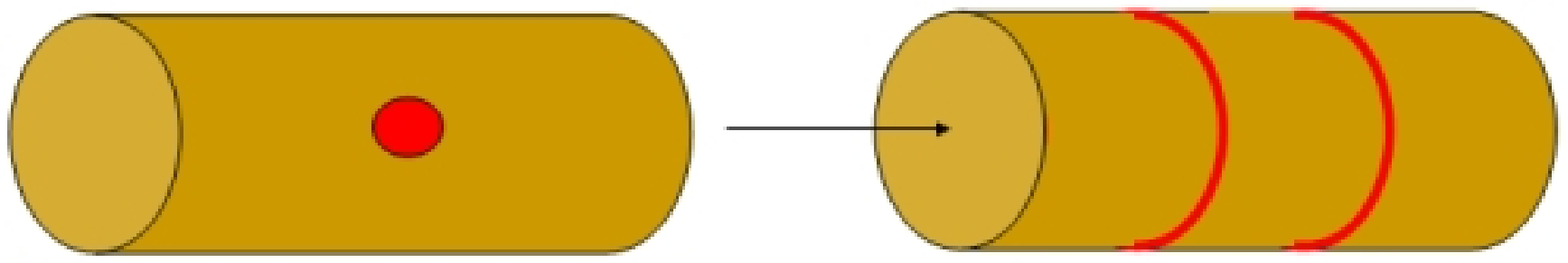,width=15cm}{A cartoon of calorons. As the lump grows, it splits into two domain walls.}{}

\para
The isolated vacua in our theory guarantee the presence of domain walls. These are BPS
and satisfy the Bogomolnyi equation
\be 2\pi H(\phi)\,\partial_1\phi =  \partial_2\sigma -\theta\label{bogwall}\ee
Typically, supersymmetric theories with two vacua have a domain wall which is BPS and interpolates
from, say, the first vacuum to the second. If we wish to go back the other way, from the second
vacuum to the first, the domain wall is anti-BPS. However, in the present situation both of these
walls can be BPS \cite{me,bruck1}. This is achieved by allowing $\sigma$ to vary along the circle, so
that the right-hand side of \eqn{bogwall} is $>0$ for the
first wall, but $<0$ for the second.  The reason that these two walls don't annihilate each
other is because the whole configuration carries the topological charge of the lump.

\para
These two domain walls form the calorons of the ${\bf CP}^1$ sigma model \cite{me,bruck1}.
In the presence of the $\theta$ term, the lump solution decomposes into two domain wall
strings as shown in the figure. This process is entirely analogous
to the caloron-monopoles appearing in Yang-Mills theories \cite{leeyi,vanbaal}. For the
${\bf CP}^{N-1}$ sigma-model, the lump decomposes into $N$ calorons.

\para
The above discussion reveals how calorons are related to the hypermultiplet partons of the microscopic
theory. Clearly they are not the same objects: the calorons are strings on ${\bf R}^2\times {\bf S}^1$, while the hypermultiplets are point-like excitations. Instead, the partons share a greater kinship
with the individual vacua rather than the domain walls since, in each vacuum, a
different hypermultiplet has condensed.

\section{Partons in the ${\bf CP}^{N-1}$ Sigma Model}
\label{cpn}

In this section we describe the $N$ partons which lie inside the soliton in the ${\bf CP}^{N-1}$
sigma model.
We will find that once again we can recover much of the lost information about the UV degrees
of freedom through a careful study of the soliton. We start by describing the UV completion of the
sigma model.

\subsection{A Quiver Gauge Theory}

\EPSFIGURE{quiver.eps,height=105pt}{}
We will construct the ${\bf CP}^{N-1}$ sigma model as the Coulomb branch of the quiver gauge
theory shown in the figure. This gauge theory  has $U(1)^N$ gauge group with
$N$ hypermultiplets, $Q_i$, $i=1,\ldots,N$. The $i^{\rm th}$ hypermultiplet has
charge $(+1,-1)$ under $U(1)_i\times U(1)_{i+1}$, where we identify $U(1)_{N+1}\equiv U(1)_1$.
For simplicity, we choose to assign each gauge group the same coupling constant $e^2$.
The overall diagonal $U(1)\in U(1)^N$ is free. Once this is removed, the $N=2$ theory coincides
with that described in Section 2.

\para
The theory has a single global symmetry $U(1)_F$, under which each hypermultiplet has charge $+1$.
By weakly gauging this symmetry, we can introduce a triplet of mass parameters, $\vec{m}$, for
the hypermultiplets. These also get masses from their coupling to vector multiplets, so the
final mass for the $i^{\rm th}$ multiplet is given by $|\vec{M}_i|$, where
\be \vec{M}_i = \vec{m}+\vphi_i-\vphi_{i-1} \label{vm}\ee
%
%
%
%
%
%
%
The vector multiplet fields are massless and the low-energy effective action is
given by a sigma model on the Coulomb branch, parameterized by the
expectation values of $\vphi_i$ and $\sigma_i$, the latter being the dual photons
defined in \eqn{dual}. Classically, the Coulomb branch is $({\bf R}^3\times {\bf S}^1)^N$.
The classical metric on the Coulomb branch is inherited from the canonical kinetic terms
of the vector multiplet fields. After integrating out the hypermultiplets, the metric
receives a correction at one-loop \cite{is} and is given by
\be ds^2 = H_{ij}\,d\vphi_i\cdot d\vphi_j + \frac{1}{4\pi}\,H^{-1}_{ij}\,(d\sigma_i + \vec{\omega}_{ik}\cdot
d\vphi_k)\,(d\sigma_j + \vec{\omega}_{jl}\cdot
d\vec{\phi}_l)\label{tstar}\ee
This is a multi-dimensional version of the Taub-NUT metric. The matrix $H_{ij}$ has components
%
%
\be
H_{ii} &=& \frac{1}{e^2} + \frac{1}{4\pi|\vec{M}_i|} + \frac{1}{4\pi |\vec{M}_{i+1}|} \label{hlots}\\
H_{ij} &=&  -\frac{1}{4\pi |\vec{M}_i|}\,\delta_{j,i-1}
-\frac{1}{4\pi|\vec{M}_{i+1}|}\,\delta_{j,i+1}\ \ \ \ \ i\neq j \nn\ee
The connection $\vec{\omega}_{ij}$ obeys
$\vec{\nabla}_iH_{ij} = \vec{\nabla}_i\times \vec{\omega}_{ij}$.  As in Section 2, non-renormalization
theorems ensure that this one-loop result is the exact description of the low-energy dynamics. Up to discrete identifications, the
metric \eqn{tstar} has the product form,
\be {\cal R}^3\times {\bf S}^1 \times {\cal M}\nn\ee
reflecting the fact that the overall, diagonal vector multiplet
is decoupled. The metric on ${\cal M}$ is hyperK\"ahler, and
closely related to the Lee-Weinberg-Yi metric for monopoles in higher rank gauge groups
\cite{lwy}.

\para
The metric on ${\cal M}$ has a $U(1)_J^{N-1}$ isometry, arising from shifts in the dual photons $\sigma^i$.
In the strong coupling limit $e^2\rightarrow \infty$, the isometry group is enhanced
to $SU(N)$. The metric
on ${\cal M}$ becomes the hyperK\"ahler metric on $T^\star{\bf CP}^{N-1}$, the cotangent
bundle of ${\bf CP}^{N-1}$.

\subsubsection*{Finding the ${\bf CP}^{N-1}$ Submanifold}

Our interest is in the solitons supported by the sigma model
on ${\cal M}$. These are BPS objects only if the vacuum state lies on the zero section of
$T^\star {\bf CP}^{N-1}$, which we will again refer to as the ``bolt". We now describe the bolt
in more detail.

\para
We take the bare mass parameter in the metric to lie along $\vec{m}=(m,0,0)$ with $m>0$.
The bolt sits within the submanifold in which $\vphi_i = (\phi_i,0,0)$. The masses \eqn{vm} then take
the form $\vec{M}_i=(M_i,0,0)$ with
\be M_i= m+\phi_i-\phi_{i-1}\nn\ee
The requirement that we lie on the bolt is simply $M_i>0$.

\para
\newcommand{\hphi}{\hat{\phi}}
Our next goal is to remove the overall free motion, parameterized  by $\sum_i \phi_i$ and $\sum_i\sigma_i$,
leaving only the $2(N-1)$ interacting fields. To this end, we define
\be \hat{\phi}_J = \phi_J - \phi_N -\left(\frac{N}{2}-J\right)m\ \ \ \ \ \ \ J=1,\ldots, N-1\label{hat}\ee
There is a similar transformation for the $\sigma_i$ variables. It is best described by first
introducing the relative field strengths,
\be \hat{F}_{\mu\nu}^J=F_{\mu\nu}^J-F_{\mu\nu}^N \ \ \ \ \ \ \ J=1,\ldots, N-1
\label{newf}\ee
The $\hat{\sigma}^I$ fields are then defined as the dual variables
\be 2\pi \hat{H}_{IJ}\hat{F}_{\mu\nu}^I = \epsilon_{\mu\nu\rho}\partial_\rho \hat{\sigma}_I
\label{ddual}\ee
In the variables $\hat{\phi}_I$ and $\hat{\sigma}_I$, the metric on the bolt can be written as
\be ds^2 = \hat{H}_{IJ}\,d\hat{\phi}_I d\hphi_J + \frac{1}{4\pi} \hat{H}^{-1}_{IJ}\, d\hat{\sigma}_I d\hat{\sigma}_J\label{cpnmet}\ee
with $I,J=1,\ldots N-1$ and the components of $\hat{H}_{IJ}$ given by
\be \hat{H}_{II} &=& \frac{N-1}{e^2N}+\frac{1}{4\pi M_I} + \frac{1}{4\pi M_{I+1}} \nn\\
\hat{H}_{IJ} &=& -\frac{1}{e^2N}-\frac{1}{4\pi M_{I+1}}\,\delta_{J,I+1} - \frac{1}{4\pi M_I}\,\delta_{J,I-1}
\label{hhat}\ee
where this notation means that
each non-diagonal element of the matrix $\hat{H}$ contains the constant piece $1/e^2N$.
The masses, $M_I$, now read
\be M_1 &=& \hphi_1 + \frac{Nm}{2}\nn\\
M_J &=& \hphi_J-\hphi_{J-1}\ \ \ \ \ J=2,\ldots,N-1\nn\\
M_N &=& \frac{Nm}{2} - \hphi_{N-1}\nn\ee
%
%
and the requirement that $M_I>0$ becomes,
\be -\frac{Nm}{2}\, \leq\, \hat{\phi}_1\, \leq\, \hat{\phi}_2\, \leq\ \ldots\ \leq\, \hat{\phi}_{N-1}\, \leq \, \frac{Nm}{2}\nn\ee
In the limit $e^2\rightarrow\infty$, equations \eqn{cpnmet} and \eqn{hhat} simply give the Fubini-Study metric on ${\bf CP}^{N-1}$ with $SU(N)$ isometry, written in toric coordinates. In contrast, for finite
$e^2$, these equations define a squashed metric on ${\bf CP}^{N-1}$ with
only $U(1)^{N-1}$ isometry.

\subsubsection*{The Example of ${\bf CP}^2$}

\EPSFIGURE{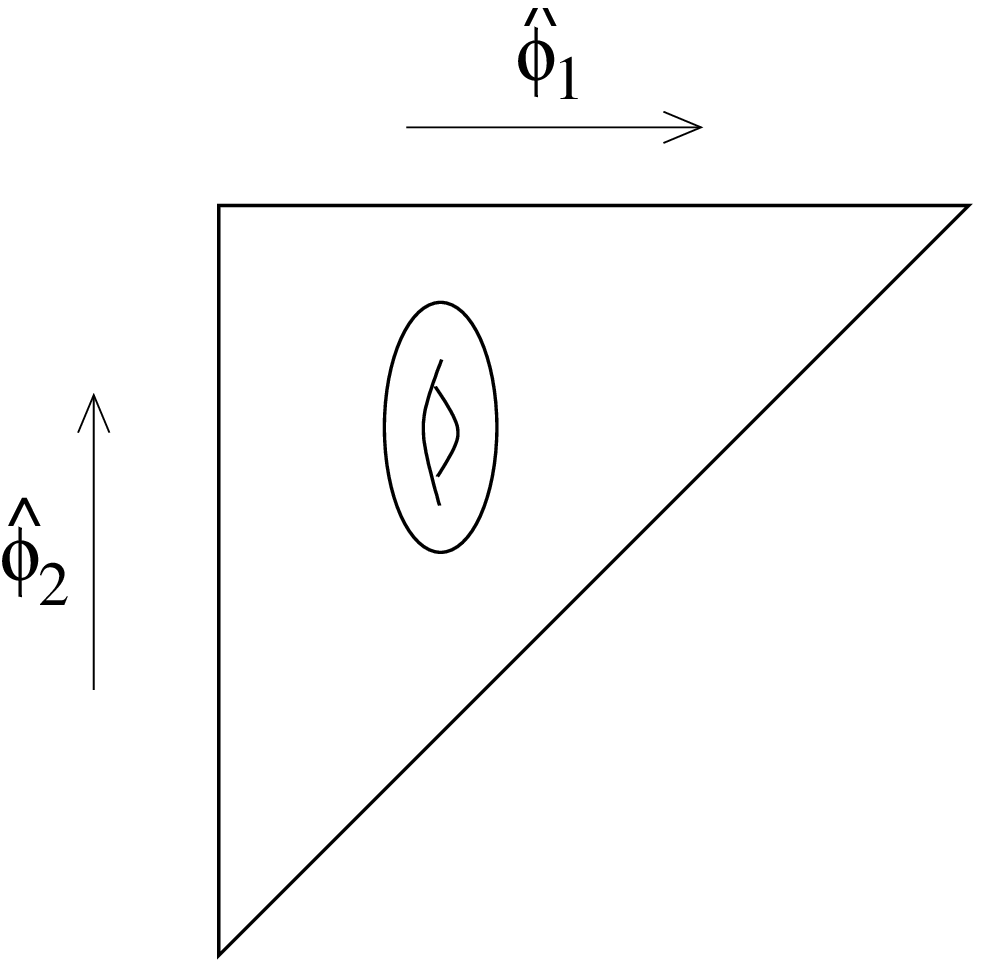,height=100pt}{}
The toric diagram for  ${\bf CP}^2$ is shown in the figure. The triangle is the region $M_i\geq 0$
for $i=1,2,3$, plotted in the $\hat{\phi}_1$ and $\hat{\phi}_2$ plane.
On each side of the triangle, one of the $M_i$ is zero, ensuring that one of the hypermultiplets
becomes massless. The left-hand edge corresponds to $M_1=0$, the upper edge to $M_3=0$, and the
diagonal to $M_2=0$.

\para
In the interior of the triangle, the two dual photons $\hat{\sigma}_1$ and $\hat{\sigma}_2$
form a torus ${\bf T}^2$, as shown in the figure. On each of the edges, one of the cycles
of the torus degenerates as dictated by the metric \eqn{cpnmet}: on the left-hand edge
$\hat{\sigma}_1$ degenerates, on the upper edge it is $\hat{\sigma}_2$, and a linear combination of
these on the diagonal.

\subsubsection*{Homogeneous Coordinates}

Before we move on to describe the solitons, it will also prove useful to describe the relationship
between our toric coordinates and the more familiar homogeneous coordinates. Each point in
${\bf CP}^{N-1}$ corresponds to an equivalence class $[f_i]$ of complex $N$-vectors $f_i \in {\bf C}$ for $i=1,\ldots N$. Two vectors $f_i$ and $\tilde{f}_i$ are equivalent if $f_i = w\tilde{f}_i$ for some complex $w\neq 0$. The relationship to toric coordinates is given by
\be \hat{\phi}^I=Nm\left(\frac{\sum_{i=1}^I |f_i|^2}{\sum_{j=1}^N|f_j|^2}-\frac{1}{2}
\right)\ \ \ {\rm and}\ \ \ e^{i\hat{\sigma}_I}=\frac{f_I}{|f_{I}|}\,\frac{|f_{I+1}|}{f_{I+1}}
\label{phif}\ee
To compare with our notation for ${\bf CP}^1$, the complex coordinate on the Riemann sphere
is given by $R=f_1/f_2$.

\subsection{Partons and Solitons}

The sigma-model on the Coulomb branch once again enjoys the presence of a soliton. The  Bogomolnyi equations
are now given by,
\be 2\pi \hat{H}_{IJ} \partial_\alpha \hat{\phi}_J = \epsilon_{\alpha\beta}\partial_\beta \hat{\sigma}_I
\label{solvethis}\ee
and a soliton with winding number $k=1$ has mass
\be M_{\rm lump} = Nm\nn\ee
A single soliton has $2N$ collective coordinates, decomposing as two center of mass coordinates, a scale size
and $2N-3$ orientation modes. These latter govern a choice of a based ${\bf CP}^1$ submanifold inside
${\bf CP}^{N-1}$.

\para
Looking to our gauge theory, there is again a unique BPS candidate for this lump: it is the gauge
invariant operator $Q_1Q_2\ldots Q_N$ constructed from a string of hypermultiplets. This object is
constructed from the $N$ links of the  quiver diagram. It carries
flavor charge $+N$, and has mass equal to that of the soliton. Moreover, it is BPS on the same locus
as the soliton.
We now show how to reconstruct this information, together with the quantum numbers of the parton,
from a study of the solitons themselves.

\subsubsection*{Deforming the ${\bf CP}^{N-1}$ Sigma Model}

Just as we saw for ${\bf CP}^1$ lumps, deforming the target space again
causes the soliton to decompose into its partonic constituents. In the
case of ${\bf CP}^{N-1}$, the target space is squashed through the
addition of the gauge coupling constant $e^2$ as in
\eqn{hhat}. The index theorem guarantees that the number of collective coordinate
of a single soliton remains $2N$ after this deformation. However, these collective
coordinates are no longer associated to Goldstone modes. Instead, they now dictate
the positions of $N$ partons.
\EPSFIGURE{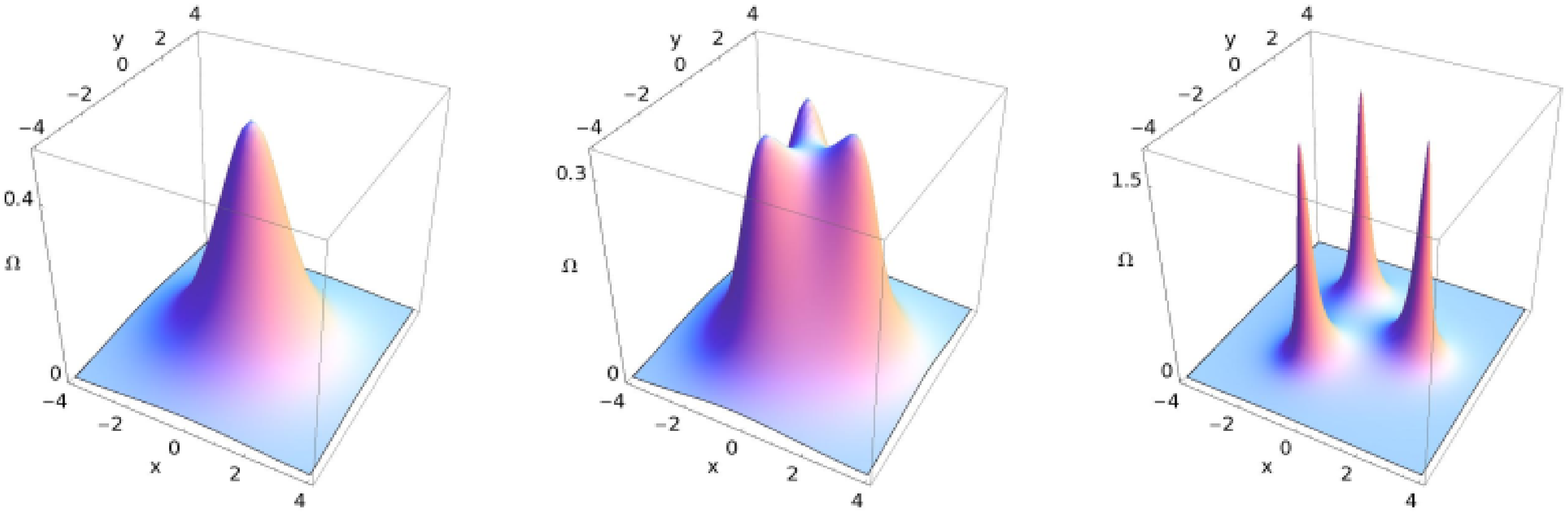,width=15cm}
{Energy density for a single ${\bf CP}^2$ soliton with $m/e^2 = 0$, $1$, $3$.}

\para
To illustrate this, in Figure 8, we plot the energy density (or equivalently, the topological
charge density) for a BPS soliton (i.e. solving \eqn{solvethis}), of winding number one, with
the target space given by a squashed ${\bf CP}^2$.
To construct these plots, we first need to choose a vacuum:
we have picked $\phi_i=0$ which ensures that all partons have equal mass.
The first part of the figure shows the lump solution for a round target
space, with $m/e^2=0$. The soliton is a smooth lump of size $\rho$ with no evidence of partonic structure.
Subsequent plots show the soliton solution for the deformed target space. As $m/e^2$ increases,
the topological charge becomes concentrated around the three points where each $M_i$ vanishes, revealing
the partonic nature of the object.

\para
In Figure 9, we plot the profile of a single $k=1$ soliton in the deformed target space with $m/e^2\approx 3$.
The overall scale size, $\rho$, of the soliton is kept fixed, while the orientation modes are changed.
The figure shows clearly that these orientation modes govern the relative positions of the partons.

\para
Note that our partons in the ${\bf CP}^{N-1}$ sigma model are {\it not} merons for $N\geq 3$. The
merons are always associated to topological charge $+1/2$, while our partons carry charge $1/N$.

\subsubsection*{Collective Coordinates and Parton Positions}

While looking at the deformed sigma-model provides the most direct way to see the partonic
nature of the soliton, is again possible to see evidence of the partons even when $e^2\rightarrow
\infty$. First, let us look at the explicit solutions in this limit.

\EPSFIGURE{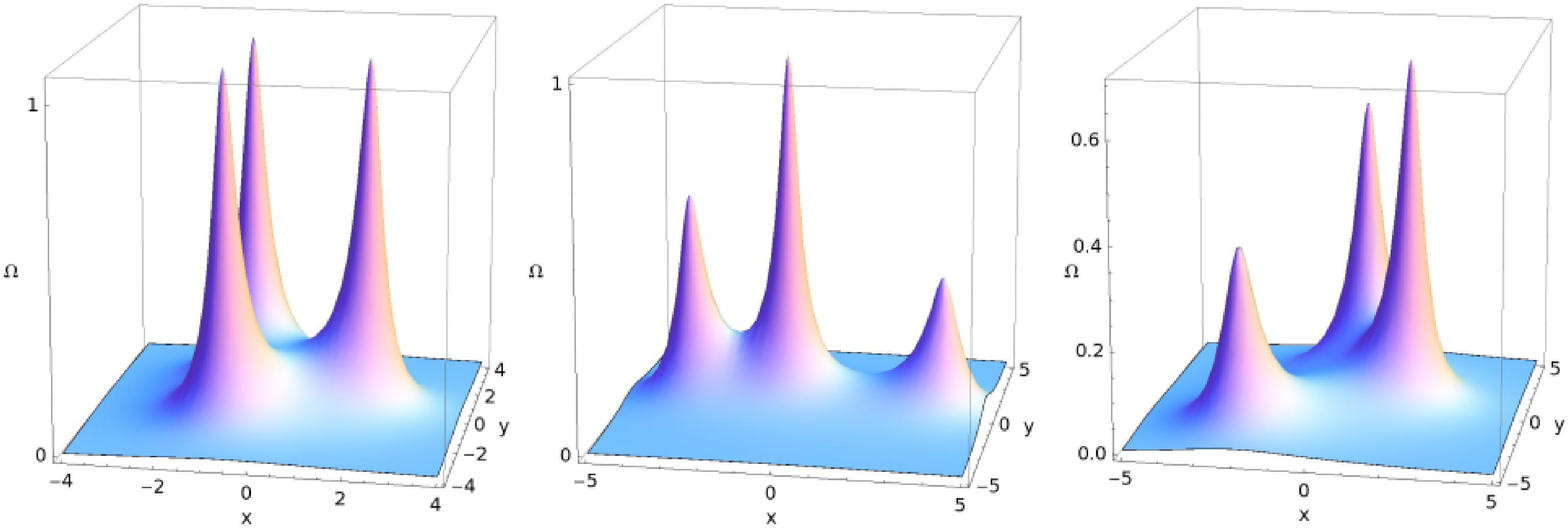,width=15cm}
{Energy density for a single ${\bf CP}^2$ soliton with fixed ``scale size" and
varying ``orientation".}

\para
The soliton with winding $k$ has $2Nk$ collective coordinates. For well separated solitons, these decompose into a position, a scale size and $2N-3$ orientation modes for each lump. However, it is
well known that the most general soliton solution is given by specifying $k$ sets of $N$
points, $\{z^1_n\}\,\ldots\{z^N_n\}$ with $n=1,\ldots, k$. In the
variables $f_i$, the soliton solution is given by
\be f_i = \prod_{n=1}^k (z-z^i_n)\label{fsol}\ee

This solution includes an implicit choice of vacuum at infinity. Examining \eqn{phif}
and \eqn{hat}, we see that this choice is
the symmetric vacuum $\phi_i=0$, in which each parton has mass $M_i=m$. In terms of
$\hat{\phi}$, this vacuum looks a little less natural: it is $\hat{\phi}_J=m(J-N/2)$.
This is the same vacuum that we picked when plotting Figures 8 and 9.

\para
It is natural to conjecture that the  points $\{z_n^i\}$ correspond to the positions of the $kN$
underlying partons.  To see that this is indeed the case, we translate the soliton
solution into toric variables. Tracing through the various definitions,
we see that the point in space $f_i=0$ corresponds to a point where the mass of the $i^{\rm th}$
hypermultiplet vanishes: $M_i=0$. This is identified as the location of the parton.

%
%

%
\EPSFIGURE{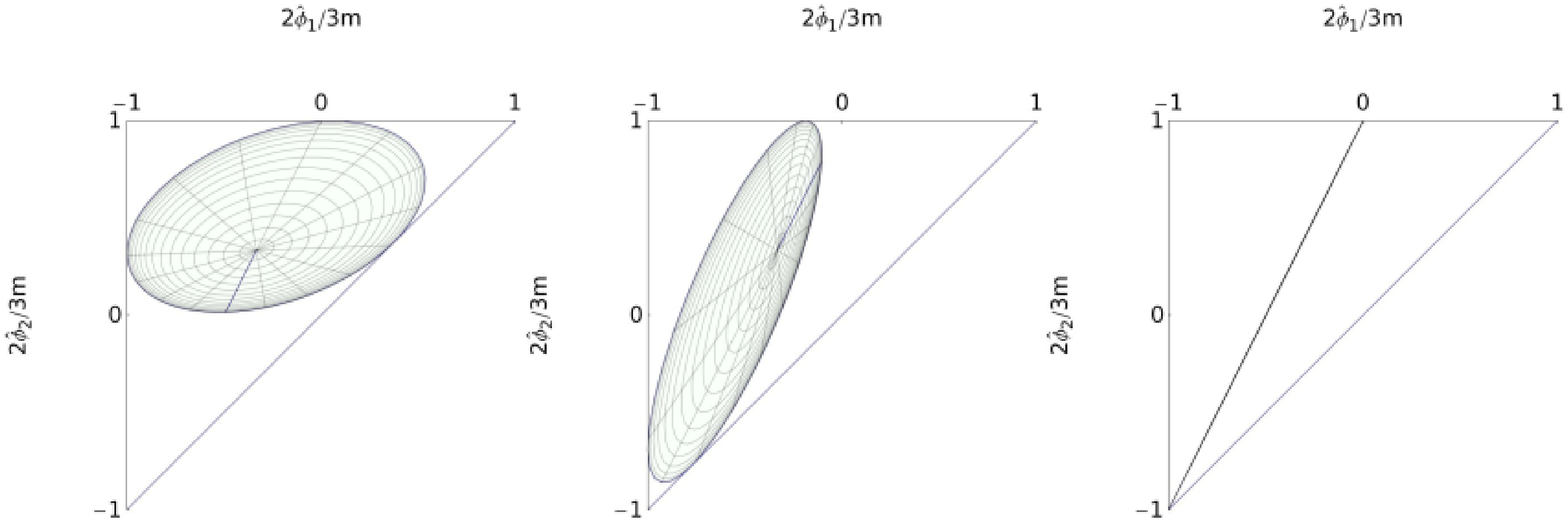,width=15cm}{Toric diagrams for ${\bf CP}^2$ solitons with different parton positions.}

\para
It is useful to illustrate these points with the example of ${\bf CP}^2$. The vacuum
$\hat{\phi}_1=-\hat{\phi}_2=-m/2$ sits firmly in the middle of the toric triangle.
The images of different lump solutions, corresponding to different ${\bf CP}^1$ submanifolds,
are shown in figure 10. In each case, the lump image touches each side of the toric diagram
at one point: these points are the images of the parton positions. The boundary of the lump image is the image of the unique circle passing through the three parton positions; the interior and exterior of this circle each map bijectively to the interior of the lump image.  The centre of the circle and the point at infinity both map to the vacuum.  In the case where the three parton positions lie on a straight line, the boundary of the lump image is the image of this line, and includes the vacuum.  Then the two half-planes on either side of the line each map to the interior of the lump image.  We can also ask what happens as two
partons approach each other. In this limit, the ${\bf CP}^1$ submanifold touches one corner
of the toric diagram, as shown in the third part of the figure.

\subsubsection*{Parton Quantum Numbers}

The parton $Q_i$ in the microscopic model carries charges $(+1,-1)$ under $U(1)_i \times U(1)_{i+1}$, for each $i=1,\ldots,N$, with $U(1)_{N+1} \equiv U(1)_1$.  We will now show that it is possible to reconstruct this pattern of electric charges of the partons from the lump solutions.

\para
We again proceed by rewriting the soliton equations in terms of dual variables by using the duality transformation \eqn{ddual}.
In these variables, the soliton equation  \eqn{solvethis} relates each electric field to the variation of the corresponding $\hat{\phi}^I$:
\be
\hat{F}_{0\alpha}^I = \partial_\alpha \hat{\phi}_I.
\label{170D1}
\ee
The electric field $\hat{F}_{0\alpha}^I$ is then divergence free except at points where one of the $\hat{\sigma}_J$ is ill-defined.  From \eqn{phif} and \eqn{fsol}, we note that these points are at $z=z_n^i$.  Each $\hat{\sigma}_J$ has non-trivial winding around two of these points.  In particular, $\hat{\sigma}_J$ increases by $2\pi$ if we complete an anticlockwise circuit around $z_n^J$, and $\hat{\sigma}_J$ decreases by $2\pi$ if we complete an anticlockwise circuit around $z_n^{J+1}$.  Following the same arguments as in section \ref{2.3}, we deduce that
\be
\partial_\alpha (\hat{H}_{IJ} \hat{F}_{0\alpha}^J) = \sum_{n=1}^k \left[ \delta(z-z_n^I) - \delta(z-z_n^{I+1})\right].
\label{almost}
\ee
Just as in the ${\bf CP}^1$ case, for given $\{ z_n^i \}$ these equations  have a unique solution which specifies a point on the soliton moduli space.

\para
Equation \eqn{almost} shows that each of the $N$ partons sources two neighbouring gauge
fields, with the exception of the first and last partons, living at positions $z_n^1$ and $z_n^N$. These
appear to be charged under just a single gauge field. To reconstruct the full quiver diagram shown
in Figure 1, it is simplest to now put back the neutral, decoupled gauge field and work with $F^i_{\mu\nu}$
with $i=1,\ldots, N$. The relationship between these $N$ gauge fields and the $(N-1)$ fields $\hat{F}_{\mu\nu}^I$ is given in \eqn{newf}. In terms of the more symmetric field strengths $F^i_{\mu\nu}$,
the source equation \eqn{almost} reads
\be
\partial_\alpha ({H}_{ij} {F}_{0\alpha}^j) = \sum_{n=1}^k
\left[ \delta(z-z_n^i) - \delta(z-z_n^{i+1})\right].
\nn\ee
where $z_n^{N+1}\equiv z_n^1$. This equation is important. It shows that the structure of
the soliton captures the quantum numbers of the partons in the UV theory. The soliton is composed
of $N$ partons, with the $i^{\rm th}$ parton carrying charge $(+1,-1)$ under
$U(1)_i\times U(1)_{i+1}$.

\subsubsection*{The Force Between Solitons}

In the case of solitons in the ${\bf CP}^1$ sigma model, we saw that the parton quantum numbers
could also be determined by the force between a soliton and anti-soliton, which coincides with
that between two dipoles. We now ask whether this property extends to the solitons of ${\bf CP}^{N-1}$.
Can we interpret the soliton anti-soliton force as the sum of dipoles with charges dictated by
the quiver? The answer appears to be no.

\para
The computation of the force between a soliton and anti-soliton in the ${\bf CP}^{N-1}$ sigma
model was performed many years ago in \cite{bard}. We'll denote the collective coordinates of
the instanton by $\{z^i\}$ as in \eqn{fsol}, while those of the anti-instanton are $\{y^i\}$.
If the two objects are separated by
distance $r$, the interaction potential is given by
\be V_{\rm int} = -\frac{4m}{r^2}\left[\sum_{i=1}^N z_iy_i-\frac{1}{N}\left(\sum_{i=1}^N z_i\right)
\left(\sum_{j=1}^Ny_i\right)\right] + {\rm h.c.}\nn\ee
This does not capture the key features of the quiver diagram. In particular, this force treats
all partons on the same footing. The $i^{\rm th}$
parton, at position $z^i$, interacts with all the anti-partons rather than just the $i^{\rm th}$
and $(i+1)^{\rm th}$ as naively suggested by the classical quiver diagram. It appears that the
filter of renormalization group flow is simply too strong and this low-energy force computation
too myopic to determine the partonic quantum numbers. Thankfully the dual Bogomolnyi equation
described above does the job for us.

\section{What Does This Tell Us About Yang-Mills Instantons?}

In $d=4+1$ spacetime dimensions, Yang-Mills theories are non-renormalizable. Arguments involving supersymmetry and string theory show that these theories have a well-defined
ultra-violet completion when equipped with 8 or 16 supercharges \cite{s5d}. Yet little
is known about the properties of the UV degrees of freedom.

\para
The story is especially interesting for the theory with 16 supercharges which has a UV
fixed point governed by the $(2,0)$ superconformal theory in $d=5+1$ dimensions. The
$(2,0)$ theory arises as the low-energy limit of $N$ M5-branes and, famously, has a
number of degrees of freedom that scales as $N^3$ \cite{klebanov}. Understanding the
kind of mathematical structure that gives rise to this  $N^3$ scaling remains an
important challenge.

\para
When
compactified on a circle of radius $R$, only $\sim N^2$ degrees of freedom remain massless
and, at long distances, the $(2,0)$ theory reduces to $5d$, maximally supersymmetric
$U(N)$ Yang-Mills with  gauge coupling $g^2 = 8\pi^2 R$. Instantons in this theory, obeying
$F={}^\star F$,  are BPS particles and are identified
with the Kaluza-Klein (KK) modes coming from six dimensions \cite{rozali},
\be M_{\rm inst} = \frac{8\pi^2}{g^2} = \frac{1}{R} = M_{KK}\nn\ee

These instantons come with a puzzle. Upon quantization, the scaling mode $\rho\in {\bf R}^+$
of the instanton gives rise to a continuous spectrum above $M_{\rm inst}$. This is odd
behaviour for a one-particle state in a quantum field theory. We propose that this continuous
spectrum arises because the instanton should be interpreted as an $N$ particle state.
Moreover, motivated by the similarity with the sigma-model described in the previous sections,
we conjecture that the $N$ partons inside the instanton are related to the UV degrees of freedom
which complete the Yang-Mills theory at high energies.

\para
Let us start by providing circumstantial evidence for this proposal. First we can ask where the
crossover from $\sim N^2$ to $\sim N^3$ degrees of freedom occurs. This was studied in \cite{imsy}
using supergravity techniques where it was shown that the transition happens at temperature
\be T \sim \frac{8\pi^2}{g^2N}\nn\ee
Indeed, this had to be the case: the theory is strongly coupled at energies $E \sim 1/g^2N$ and
this is where the new degrees of freedom must kick in. This mass scale points firmly at the
existence of instantonic partons.

\para
The existence of modes carrying fractional KK momentum is familiar in compactified Yang-Mills theories, where they arise in the  presence of a Wilson line. Such modes occur whenever $N$ branes wrapped on
the circle combine to form a single ``long" brane wrapped $N$ times. Behaviour of this type was
important in the original work on black hole entropy counting \cite{sumit,sm}.

\para
It is also worth mentioning some further numerological evidence for the partonic interpretation
of instantons. The $N^3$ scaling for  M5-branes can be refined by an
anomaly computation \cite{hmm}
whose coefficient provides the subleading term in the number of degrees of
freedom on $N$ M5-branes: $c(su(N)) = N^3-N$. A generalization
of this formula to other $G=\,$ADE theories was conjectured by Intriligator to be \cite{ken},
\be c(G) = C_2(G)\,|G|\nn\ee
where $C_2(G)$ is the dual Coxeter number (normalized such that $C_2(su(N))=N)$
and $|G|$ is the dimension
of the group. This fits nicely with our partonic interpretation of instantons since the dimension
of the moduli space of a single instanton is $4C_2(G)$, implying the existence of $C_2(G)$ partons
in general. The presence of $|G|$ in the anomaly coefficient is perhaps hinting that
each of these partons transforms in the adjoint of the gauge group $G$.

\para
In the case of sigma-model solitons, we have seen above that a detailed study allows us to reconstruct
properties of the high-energy theory. Can we do something similar for Yang-Mills instantons?
We hope to return to this question in future work. Here we limit ourselves to a few simple
observations and speculations.
Firstly, the instanton solution has only magnetic components of the five-dimensional
gauge field turned on. Yet, in five dimensions, magnetic charge is naturally carried by
string-like objects. So perhaps each parton is itself a loop of string. Indeed, the caloron
picture \cite{leeyi,vanbaal} reveals $N$ strings inside the instanton but, as we described
in Section 2, in the case of sigma-model lumps calorons were not directly related to
partons. Strings were also found lurking inside instantons in \cite{sk1,sk2} in the
context of dyonic instantons \cite{frodo}.

\para
In the context of the sigma-model, the force between a lump and anti-lump spectacularly
revealed the partonic quantum numbers for ${\bf CP}^1$, but proved more myopic in the case of
${\bf CP}^{N-1}$. For Yang-Mills instantons, the force was computed in \cite{callan}. An
instanton of size $\rho$ and an anti-instanton of size $\bar{\rho}$, separated by a distance
$r\gg \rho,\bar{\rho}$  feel the attractive potential
\be V = -\frac{32\pi^2}{g^2}\frac{\rho^2\bar{\rho}^2}{r^4}\,C_{ab}\,\bar{\eta}^a_{\mu\nu}
\eta^b_{\mu\lambda}\hat{r}_\nu\hat{r}_\lambda\nn\ee
The fact that the force is quadratic in $\rho$, rather than linear, is again
indicative of loop-like objects as befits a magnetic dipole.
Here $C_{ab}$ describes the relative orientation of the two instantons within the gauge
group. Aligned instantons, with $C_{ab}=\delta_{ab}$ feel the maximum force. However,
in contrast to the situation with sigma-model lumps, instantons
can hide from each. If they sit in commuting $SU(2)$ factors in the gauge group, the instanton and anti-instanton
feel no force. (A similar phenomenon is not allowed
in the case of the sigma-model because both lump and anti-lump
solutions are required to asymptote to the same vacuum). Needless to say, it would be very
interesting to interpret the instanton force formula in terms of partons.

\para
Finally, perhaps the most important question is to determine the confinement mechanism
that binds the
partons inside the instanton yet allows them to move freely.
In the case of the sigma-model lump this arose due to the
log-divergent energy arising from the long-range fields of the parton. However, as discussed above, this
divergence reveals itself in the moduli space metric for lumps. There is no hint of such a
divergence in the moduli space metric for instantons, suggesting that the confinement mechanism
is something different in this case.
In particular, this means that the merons discussed in \cite{callan} are
not the partons of interest: as well as having topological charge $1/2$ instead of $1/N$, they have $F_{\mu\nu} \sim 1/r^2$ giving rise to a log-divergent energy.  It appears that the confinement
mechanism at play inside Yang-Mills instantons is somewhat more subtle. Perhaps the deconstruction of
the $(2,0)$ theories  presented in \cite{nima,gm} can shed light on this issue.

\section*{Acknowledgement}

Our thanks to Nima Arkani-Hamed, Adi Armoni, Nick Dorey, Tim Hollowood, Prem Kumar, Walter Vinci for useful discussions. 
BC is supported by an STFC studentship.  DT is supported by the Royal Society. Mathematica notebooks
for all figures presented in this paper can be downloaded from:
\begin{center}
http://www.damtp.cam.ac.uk/user/tong/parton.html
\end{center}

\end{document}